\documentclass[manuscript, screen]{acmart}

\AtBeginDocument{%
  }
  \acmConference[CHI '26]{CHI Conference on Human Factors in Computing Systems}{April 13--17,
  2026}{Barcelona, Spain}
\acmBooktitle{CHI '26: Proceedings of the CHI Conference on Human Factors in Computing Systems (CHI '26), April 13--17,
  2026, Barcelona, Spain}

\usepackage{float}
\usepackage{multirow}   
\usepackage{booktabs}   
\begin{document}

\title{“Koyi Sawaal Nahi Hai”: Reimagining Maternal Health Chatbots for Collective, Culturally Grounded Care}


\author{Imaan Hameed}
\affiliation{%
  \institution{Lahore University of Management Sciences (LUMS)}
  \city{Lahore}
  \country{Pakistan}}
\email{24100093@lums.edu.pk}

\author{Huma Umar}
\affiliation{%
  \institution{Lahore University of Management Sciences (LUMS)}
  \city{Lahore}
  \country{Pakistan}}
\email{huma.umar@lums.edu.pk}

\author{Fozia Umber}
\affiliation{
  \institution{Lahore Medical and Dental College}
  \city{Lahore}
  \country{Pakistan}}
\email{fozia.umber@ubas.edu.pk}

\author{Maryam Mustafa}
\affiliation{%
  \institution{Lahore University of Management Sciences (LUMS)}
  \city{Lahore}
  \country{Pakistan}}
\email{maryam.mustafa@lums.edu.pk}


\begin{abstract}
In recent years, LLM-based maternal health chatbots have been widely deployed in low-resource settings, but they often ignore real-world contexts where women may not own phones, have limited literacy, and share decision-making within families. Through the deployment of a WhatsApp-based maternal health chatbot with 48 pregnant women in Lahore, Pakistan, we examine barriers to use in populations where phones are shared, decision-making is collective, and literacy varies. We complement this with focus group discussions with obstetric clinicians. Our findings reveal how adoption is shaped by proxy consent and family mediation, intermittent phone access, silence around asking questions, infrastructural breakdowns, and contested authority. We frame barriers to non-use as culturally conditioned rather than individual choices, and introduce the Relational Chatbot Design Grammar (RCDG): four commitments that enable mediated decision-making, recognize silence as engagement, support episodic use, and treat fragility as baseline to reorient maternal health chatbots toward culturally grounded, collective care.

\end{abstract}





\maketitle

\section{Introduction}

Pregnancy care in low-resource health systems is often precarious~\cite{who2024inequalitymonitor,who2015trendsmaternal}. In Pakistan, where maternal mortality remains high, public hospitals contend with overcrowding, limited staffing, and fragile bureaucratic structures. Digital health interventions—particularly conversational agents—are increasingly promoted as cost-effective~\cite{stanfordcdh2023whitepaper,almalki2020covidchatbots,wilson2022publichealthchatbots, hussain2025revolutionize} ways to expand access to information and ease provider workloads. However,  underlying assumptions of privacy, phone ownership, literacy and autonomy do not hold true in low-resources, conservative contexts. This assumption contrasts sharply with pregnancy in South Asia, where health-related decisions are typically collective, phones are shared, and care-seeking during pregnancy is shaped as much by endurance and negotiation as by individual autonomy. 

HCI research has examined both the promises and limitations of health chatbots. Conversational agents are now widely deployed, with the rise of large language models accelerating their spread across domains~\cite{singhal2023clinicalknowledge,zhou2023surveyllmmedicine,who2024lmmguidance}. In healthcare, chatbots are framed as scalable tools for patient education, triage, and treatment adherence~\cite{parmar2022reviewapps,wilson2022publichealthchatbots,montenegro2022pregnancychatbot}. Governments and NGOs have introduced them through SMS and WhatsApp, including in maternal health, where they are promoted as addressing gaps in information and continuity of care~\cite{mcmahon2023nursenisa,defilippo2023chatbotgender}. 

Yet, adoption remains uneven. Pilot studies often show strong initial interest but rapid drop-off~\cite{stowell2018mhealthvulnerable,coleman2023reconsidering}, while large-scale deployments such as ASHAbot demonstrate technical feasibility~\cite{khullar2025ashabot}. Critical work has highlighted persistent challenges of trust, interpretability, and cultural fit. Studies from the Global South to reveal that surface-level localization—such as simple translation—is rarely sufficient\cite{khullar2025ashabot,deva2025culturalsensitivity,rahman2021adolescentbot,yadav2019feedpal,wang2022snehai}. Instead, adoption of technologies is shaped by broader social and institutional dynamics. Broader surveys echo these patterns, linking non-use not only to interface design but also to uncertainty, mistrust, and systemic barriers~\cite{baumer2015nonuse,satchell2009beyonduser}. These studies highlight that silence, refusal, and abstention are analytically meaningful, but they do not extend to gendered or maternal health settings in South Asia or even the Global South more broadly. As a result, the everyday practices of silence, proxy use, and collective consent that shape care, non-use, and drop-off in these contexts remain under-theorized in chatbot design.

These findings underscore that availability alone does not ensure use; uptake remains uneven. 

In this paper, we pose the following research questions:
\begin{enumerate}
    \item How do pregnant women and their families engage with, adopt, or resist a maternal health chatbot within the routines of everyday care?
    \item How do social, medical, and systemic factors shape uptake in fragile health systems?
    \item What design opportunities and challenges arise in building maternal health chatbots that are trusted, inclusive, and safe in low-resource contexts?
\end{enumerate}

We report on the deployment of a maternal health chatbot in Lahore, Pakistan, working with 48 pregnant women at both a public and a private hospital, alongside focus groups with gynecologists. Our findings show that adoption is influenced less by interface usability and more by sociotechnical dynamics: collective consent, family mediation, shared phone use, infrastructural breakdowns, and institutional gate-keeping.

Our contributions are threefold:

\textbf{Empirical:} We provide an empirical account of how maternal health chatbot use in Lahore was shaped by collective consent, proxy mediation, silence, episodic access, and systemic fragility.

\textbf{Design:} We translate these findings into a \textit{Relational Chatbot Design Grammar (RCDG)} with four commitments—Mediated Decision-Making, Silence \& Endurance, Episodic Use, and Fragile Contexts—operationalized through layered consent, narration-to-question translation, episodic check-ins, and disruption-aware re-entry.

\textbf{Conceptual:} We reframe chatbots as relational and situated technologies, extending HCI debates on non-use by showing how silence, withholding, and discretion function as patterned practices of participation in fragile, collective care contexts.

By centering the conditions of adoption, this work broadens HCI debates on the sociotechnical lives of digital health tools, particularly within health systems in the Global South.

\section{Related Work}

Prior studies of LLMs and health chatbots emphasize technical promise but often underplay the complicated ground realities that chatbots are used in. HCI research on maternal health in LMICs has begun to surface these dynamics, while qualitative scholarship on reproductive health more broadly reveals the structural forces that continue to shape maternal care. Bringing these literatures together frames silence, family mediation, and structural instabilities not just as barriers to uptake, but as design concerns that need to be interrogated while developing chatbots.

\subsection{LLMs and Health Chatbots}
Recent work on LLMs in healthcare highlights both their promise—personalized, open-ended, multilingual support across mental health, caregiving, and public health~\cite{jo2023publichealthllm,sharma2024cognitiverestructuring,song2024typingcure,wang2021gpttherapy,yang2023interpretablemh,ma2024explorellm}—and persistent debates around safety, interpretability, and contextual fit in fragile or inequitable systems. These studies situate LLMs as transformative but also caution that their integration into care requires careful governance and cultural sensitivity. 

Reviews show that while health chatbots have often been deployed as rapid-response solutions during crises, especially COVID-19~\cite{almalki2020covidchatbots,wilson2022publichealthchatbots}, evaluations typically privilege technical throughput and information delivery. Adjacent work on large-scale counseling conversations and patient assessments demonstrates that engagement hinges less on volume than on trust cues, recognition, and relational quality, dimensions largely absent from chatbot evaluations~\cite{althoff2016counseling,sillence2007evaluateonlinehealth,sambasivan2010intermediated}. 

Evaluations of health chatbots and mHealth interventions frequently equate success with sustained engagement, reporting metrics such as log-ins, sessions, or time spent \cite{wang2022snehai,mcmahon2023nursenisa,yadav2019feedpal,rahman2021adolescentbot,montenegro2022pregnancychatbot}. While some reviews critique this focus and call for broader measures of equity and context \cite{katell2020situated,veale2017fairerml,till2023digitalhealthafrica,khullar2025capabilities}, most reported evaluations center on activity metrics (query counts, DAU/WAU, escalations), with far less attention to situated outcomes such as safe handoffs, refusal quality, equity of reach, or time-to-care~\cite{parmar2022reviewapps,wilson2022publichealthchatbots,khullar2025ashabot}. 

While LLMs demonstrate strong clinical knowledge encoding and application potential~\cite{singhal2023clinicalknowledge}, their integration into care settings also raises governance, ethical, and equity concerns~\cite{who2024lmmguidance,stanfordcdh2023whitepaper,sambasivan2010intermediated}, with surveys underscoring both promising use cases and unresolved risks across medicine~\cite{zhou2023surveyllmmedicine}. 

Much of the work on maternal and reproductive health chatbots in LMICs has framed their role narrowly as vehicles for information transfer~\cite{yadav2019feedpal,wang2022snehai,rahman2021adolescentbot,parmar2022reviewapps,wilson2022publichealthchatbots}. Studies on breastfeeding education, reproductive health case studies, and adolescent sexual health interventions illustrate this orientation. Systematic reviews similarly show that maternal health chatbots and public health conversational agents tend to privilege information provision over relational or contextual support~\cite{parmar2022reviewapps,wilson2022publichealthchatbots}. Recent work also cautions that AI chatbots, while expanding access to health information, risk amplifying misinformation and undermining trust if not carefully validated and governed~\cite{meyrowitsch2023misinfo}.

Yet, in many under-served contexts, the challenge is less about a lack of information and more about how people are able—or unable—to act on it. Health informatics interventions often falter for marginalized groups because they overemphasize individual behavior change, overlooking structural barriers and thereby risking intervention-generated inequalities~\cite{veinotetal2019,veale2017fairerml}. Care-seeking is further complicated by the coexistence of biomedical advice and long-standing cultural or traditional health practices, which are often in tension~\cite{mustafa2020patriarchy,dsanensor2025ghana}. Research on shared-phone use in South Asia shows that privacy and consent are rarely individualized but instead negotiated within households \cite{ahmed2019personalstuff,naveed2022privacyperceptions}. 

Despite extensive evaluations of health chatbots and growing work on LLMs in medicine, most studies still treat information transfer and activity metrics as the central measures of success, leaving underexplored how maternal and reproductive health chatbots should account for collective decision-making, proxy use, and structural barriers in LMICs.

\subsection{Negotiated and Collective Care in HCI}

HCI work on maternal and child health in LMICs consistently shows that clinical knowledge alone does not determine practice \cite{yadav2019feedpal,wang2022snehai,rahman2021adolescentbot}. Everyday care is negotiated across families, frontline workers such as community health workers, clinicians, and fragile institutions \cite{deva2025culturalsensitivity,manvi2025receptivity,ko2025domainexperts,mburu2018codesigning}. 

Bagalkot et al. argue that health literacy framings that dictate chatbot logics are misplaced. For them, the problem is that many women already know the recommendations but must negotiate authority with husbands, mothers-in-law, and providers. Hence digital tools should support collective negotiation and trust-building, not just information transfer \cite{bagalkot2018everydaypregnancy,bagalkot2020beyondliteracy}. A similar asset-based lens from Ghana reframes sociocultural tensions between traditional and biomedical practices as opportunities for design, recognizing elders and community leaders as resources to engage rather than barriers to overcome \cite{dsanensor2025ghana,ko2025domainexperts,wongvillacres2018collectiveaction}. 

Trust is a central trope in chatbot research. Studies link it to accuracy, transparency, and empathetic tone \cite{silva2024chatdesign,liu2018chatbots,progga2025perinataltech, harrington2023trustcomfort}. In recent LMIC deployments such as ASHABot, community health workers treated the bot as authoritative when its answers aligned with guidelines \cite{deva2025culturalsensitivity}. 

Prior digital interventions highlight these tensions: Batool et al. found SMS and voice messages improved knowledge but not hospital visits, since mobility and decisions remained with kin~\cite{batool2017maternal}. Sultana et al. show how women had to use technologies tactically within patriarchal surveillance~\cite{sultana2018patriarchal}. Verdezoto et al. \cite{verdezoto2021maintenance} highlight the invisible maintenance work CHWs perform to repair community trust in medical advice, while large-scale studies in Africa and other LMICs show how fragile infrastructures, high workloads, and inequitable health systems constrain the impact of mHealth and decision support interventions \cite{till2023digitalhealthafrica,coleman2023reconsidering}. 

Comparative studies further underscore these dynamics. A pregnancy chatbot pilot in Brazil showed potential for reducing anxiety but revealed that adoption hinged on infrastructure and integration into daily life \cite{montenegro2022pregnancychatbot}. In the U.S., Chaudhry et al. \cite{chaudhry2019pregnancyapp} demonstrated that tailoring a pregnancy app to low-income women offered gains, yet remained tied to a single-user, single-device model. Perinatal mental health research also points to parallel dynamics of trust, anonymity, and misinformation, strengthening calls for provenance of evidence and longitudinal support \cite{progga2025perinataltech}. In Pakistan, earlier ICT work illustrated both the promise and the limits of community infrastructures: technologies improved access and accountability but remained vulnerable to outages, content gaps, and socio-demographic inequalities \cite{zakar2014itwomenpk}.

Some scholarship in HCI has treated non-use as a meaningful analytic category, though almost entirely outside of gendered or maternal health contexts. Early work in Australia emphasized how non-use itself reveals important relations between people and technologies \cite{satchell2009beyonduser}, while more recent theorizing from the U.S. critiques how the category of ‘non-use’ can ultimately reinscribe use as the central analytic \cite{baumer2025sociotechnical}. Studies of AI in UX workplaces in North America likewise frame non-use as a spectrum of situated practices shaped by organizational and professional constraints \cite{cha2025ainonuse}. These seeds of a conversation on non-use, silence, refusal have not yet been extended to maternal or reproductive health chatbots in South Asia and other LMICs.

Work on repair and breakdown in HCI—largely developed through studies of infrastructures in North America—shows that technologies are inherently fragile and always in flux \cite{jackson2014rethinkingrepair,jackson2014breakdown,ismail2021frontlines}. Research on resilience in collaboration, grounded in crisis response and displacement contexts in the U.S., further demonstrates how communities sustain use through situated improvisation under conditions of disruption \cite{mark2009resilience}.

Seen through repair and resilience lenses~\cite{jackson2014rethinkingrepair,jackson2014breakdown,mark2009resilience}, recent LLM pilots in India attempt to localize within already-fragile infrastructures \cite{khullar2025ashabot,deva2025culturalsensitivity}. ASHABot integrates GPT-4 with curated knowledge bases and escalation to Auxiliary Nurse Midwives, creating a stigma-free channel for community health workers while embedding iterative safeguards and language adaptation \cite{khullar2025ashabot} but evaluated primarily via activity measures (volume, escalations) rather than situated safety or equity outcomes. Deva et al. \cite{deva2025culturalsensitivity} co-designed a culturally sensitive SRH chatbot with an empathetic gynecologist persona, where usage patterns revealed queries during “quiet hours” that intertwined medical, familial, and safety concerns. Yet even these progressive designs share assumptions that women and workers have consistent access to personal phones, stable WiFi access, and functioning clinician escalation channels.

Across diverse contexts—India, South Africa, Ghana, Brazil, Pakistan, and the U.S.—a shared pattern emerges. Digital health systems often presume literate, independent users with stable, individual phone access. In practice, women navigate pregnancy and maternal care through collective decision-making, proxy use, silence or deference, and intermittently available devices and infrastructure \cite{bagalkot2018everydaypregnancy,verdezoto2021maintenance,till2023digitalhealthafrica,mustafa2020patriarchy,coleman2023reconsidering,montenegro2022pregnancychatbot,zakar2014itwomenpk}. Despite pervasive non-use and silence in maternal care, SRH/maternal chatbots in LMICs almost never theorize these practices beyond attrition statistics \cite{satchell2009beyonduser,baumer2025sociotechnical,cha2025ainonuse}. This divergence underscores the need for rethinking maternal health chatbots not as individual knowledge tools but as social infrastructures designed for materially and culturally situated care.

While HCI research has begun to recognize negotiation, repair, and non-use in other domains, maternal and reproductive health chatbots in LMICs rarely theorize silence, refusal, or collective care as analytic categories, leaving a gap in understanding how these practices shape use and non-use.

\subsection{Designing for Reproductive Health and Structural Constraints in HCI}

For HCI, reproductive technologies like chatbots must be situated within existing analyses of what governs reproductive life. Historical work shows that other reproductive health interventions, such as family planning, for example, have been driven by state agendas and global governance, not only by women’s needs \cite{baron2008origins,bracke2021unreprorights}. For feminist HCI, this matters because it reminds us that reproductive health technologies are never neutral—they shape and reflect authority~\cite{bardzell2010feministhci,bardzell2011methodology,bellini2018femhcicommunity}. Critical work on automation argues that bots and conversational agents encode values and reproduce power relations \cite{richterich2024feministautomation,winkle2023feministhri,masure2019feministalexa}. This also builds on Ismail and Kumar’s critique of techno-optimism in the Global South \cite{ismail2021frontlines}, which governance and authority, assuming technology can “add trust” where institutions falter. These perspectives call for feminist and justice-oriented approaches to conversational design.

Phelan, Link, and Tehranifar \cite{phelan2010fundamentalcauses} describe these as “fundamental causes”—education, wealth, and kinship ties, among others—that continuously shape risk. The reproductive justice framework~\cite{ross2017reprojustice} reminds us that maternal health chatbots do more than deliver information—they operate within broader social and structural conditions that shape women’s access, autonomy, and ability to act. For example, in Egypt, family planning campaigns framed contraception as voluntary choice while furthering state rationalization \cite{ali2002planningfamily}, and in Pakistan, Islamic idioms have been mobilized both to legitimate and to resist family planning \cite{varley2012pakistan,ataullahjan2019resistance,mahmood1997populationplanning}.

Regional evidence on reproductive health more generally reflects this: contraceptive uptake correlates with education, employment, and media access, while women in Pakistan face stark inequalities in access to reproductive health care by income and literacy \cite{sharma2022contraceptiveineq,sheikh2018educatinggirls,fikree2004genderdisparity}. Despite gains in antenatal and facility-based care, mortality remains high, underscoring that information delivery alone does not resolve entrenched inequities \cite{nips2019pdhs,qadir2011genderpreference}.  Structural limitations compound these inequalities: Pakistan has just 10.8 doctors and 5.2 nurses per 10,000 people—far below the WHO’s minimum threshold of 30 per 10,000, and behind the global average of about 38 per 10,000 \cite{WHO2021workforce,PBC2025nursing}. 

Feminist political theory conceptualizes autonomy as socially embedded rather than purely individual \cite{mackenzie2000relationalautonomy}, while justice requires attention to structural inequalities and group differences \cite{young1990justice, banerjea2018liveable}. These frameworks reorient design away from individualized notions of empowerment.

Liberal approaches, such as Sen’s capability framework \cite{sen1999development}, often frame agency in terms of individual freedoms and choices. Feminist and anthropological scholarship further complicates this: Mahmood \cite{mahmood2005politicsofpiety} and White \cite{white2010fundamentalcauses} demonstrate how agency can be cultivated through virtue, devotion, and obligation, while Rowlands \cite{rowlands1997questioningempowerment} reconceptualizes empowerment as communal and collective, rather than just individual. Kabeer \cite{Kabeer1999} similarly argues that empowerment must be understood as the interplay of resources, agency, and achievements. Banerjea et al. \cite{banerjea2018liveable} build on this to show how people make life "liveable" within constraints, using tactics like discretion and endurance as everyday ways of maintaining dignity and care. 

These accounts show how the very categories of “choice” and “empowerment” that underlie information-transfer logics are themselves culturally situated. Although feminist scholarship across HCI and social sciences remind us that technologies are embedded in power, inequality, and cultural idioms, maternal health chatbots remain designed mostly through individualized notions of empowerment, leaving a gap in understanding how to build socially embedded, justice-oriented infrastructures of care in the Global South.

Existing work on LLMs, health chatbots, and maternal care in HCI highlights technical promise, negotiation, and structural constraints, but it remains fragmented—focusing on information transfer, activity metrics, or individual empowerment—without theorizing silence, non-use, and collective care as central design concerns in LMIC contexts.

\section{ Methodology}
\begin{figure}
    \centering
    \includegraphics[width=1\linewidth]{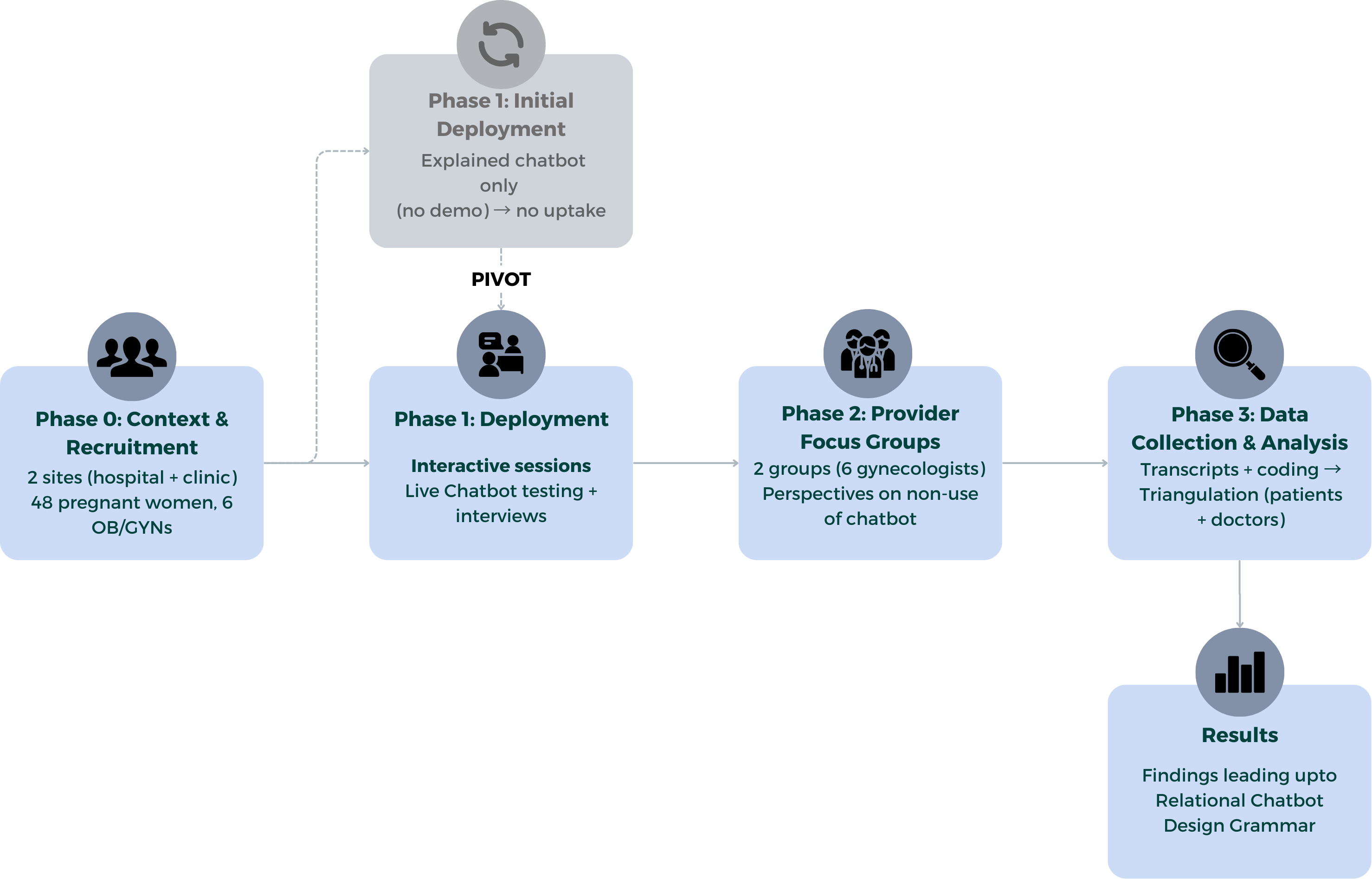}
    \caption{\textbf{Study phases.} The study unfolded across four sequential phases. \textit{Phase 0} established context and recruitment, \textit{Phase 1} piloted and then pivoted the deployment strategy, \textit{Phase 2} gathered provider perspectives, and \textit{Phase 3} synthesized data through analysis and triangulation. Solid arrows show the planned sequence of phases, while the dotted arrow marks the pivot from the initial deployment attempt.}
    \label{fig:studyphases}
\end{figure}

\subsection{Study Context}

We piloted a WhatsApp-based maternal health chatbot in Lahore, Pakistan, working primarily at a large not-for-profit teaching hospital serving predominantly low- to middle-income women, and supplemented this with a field visit to a private clinic with a dedicated obstetrics and gynecology wing. These contrasting sites offered variation in patient demographics, resources, and provider practices, allowing us to examine how adoption was shaped not only by individual and familial dynamics, but also by institutional contexts that differed in scale, infrastructure, and access.

\subsection{System Design}

The multimodal maternal health chatbot is built on OpenAI’s thread-based API architecture. The system integrates speech recognition, large language model reasoning, text-to-speech synthesis, and WhatsApp delivery into a pipeline optimized for continuous, natural interaction.

Each user is assigned a dedicated conversational thread that preserves system instructions, preferences, and complete dialogue history. Unlike stateless models, this design removes the need to resend histories with every request, reducing latency and enabling richer contextual understanding.

User inputs arrive via WhatsApp through Twilio’s API\footnote{\url{https://www.twilio.com/en-us}}, as text or audio. Audio messages are transcribed using OpenAI’s GPT-4o\footnote{\url{https://openai.com/index/hello-gpt-4o/}}, which preserves conversational nuance. The transcribed or text input is then processed by the Assistants API\footnote{\url{https://platform.openai.com/docs/assistants/migration}}, which generates a contextually grounded response using the persistent thread. Responses are returned in both text and natural voice, synthesized through ElevenLabs\footnote{\url{https://elevenlabs.io/}}, and delivered back to WhatsApp.

\subsection{Participants and Recruitment}

We engaged 48 pregnant women during antenatal visits. (Table  \ref{tab:demographics}) highlights detailed demographics: Participants were aged 20–38 (median 27). Most were housewives in joint households, with monthly incomes of PKR 30,000–100,000. Education ranged from no formal schooling to postgraduate degrees. About 70\% owned a WhatsApp-enabled phone, 20\% relied on a shared device, and 10\% had no access. Health information was typically sought from family, doctors, and social media.

We also conducted two focus groups with six gynecologists (Table \ref{tab:doctor_demographics}) from major teaching hospitals in Lahore, which helped contextualize patient experiences by highlighting clinical expectations, institutional routines, and evidence-based guidelines.

Recruitment varied across sites. At the teaching hospital, researchers approached women in waiting areas, with some referrals by doctors; at the private clinic, participants were consistently referred by clinicians after consultation. In both settings, support from hospital staff increased willingness to participate.

\begin{table}
\centering
\caption{Participant demographics ($n=48$).}
\label{tab:demographics}
\begin{tabular}{@{}lll@{}}
\toprule
\textbf{Category} & \textbf{Subcategory (range/label)} & \textbf{\%} \\
\midrule
Age & 20--24 years & 25 \\
    & 25--29 years & 30 \\
    & 30--34 years & 28 \\
    & 35--38 years & 17 \\
\midrule
Education & No formal / Primary & 12 \\
          & Lower secondary & 21 \\
          & Higher secondary & 17 \\
          & Undergraduate & 15 \\
          & Postgraduate  & 13 \\
\midrule
Employment & Housewives & 79 \\
           & Professionals & 21 \\
\midrule
Household income & 25k--50k PKR (\$90--180) & 35 \\
                 & 50k--100k PKR (\$180--360) & 40 \\
                 & 100k--200k PKR (\$360--720) & 15 \\
                 & 200k--450k PKR (\$720--1600) & 10 \\
\midrule
Family setup & Joint households & 75 \\
             & Nuclear households & 25 \\
\midrule
Pregnancy history & First pregnancy & 21 \\
                  & 2--3 pregnancies & 42 \\
                  & 4 or more pregnancies & 10 \\
\midrule
Phone access & Own smartphone & 71 \\
             & Shared phone & 21 \\
             & No smartphone / button phone only & 8 \\
\midrule
Ethnicity & Punjabi & 90 \\
          & Other (Dir, Sindh) & 10 \\
\bottomrule
\end{tabular}
\end{table}

\begin{table}
\centering
\caption{Overview of qualifications, training stage, and clinical experience of participating OB/GYN doctors.}
\label{tab:doctor_demographics}
\begin{tabular}{p{1cm} p{3.5cm} p{4cm} p{3cm}}
\toprule
\textbf{ID} & \textbf{Qualification} & \textbf{Training Stage / Role} & \textbf{Years of Experience} \\
\midrule
D1 & MBBS & PG-3 Resident & 3 years residency \\
\midrule
D2 & MBBS, MCPS & FCPS Trainee & MCPS, FCPS \\
\midrule
D3 & MBBS & PG-1 Resident & House job + 1st year PG \\
\midrule
D4 & MBBS, MRCOG Part 1 & Research Associate & 2 years \\
\midrule
D5 & MBBS, FCPS Part 1 & PG-2 Resident & 2 years \\
\midrule
D6 & MBBS & Postgraduate training & Current training \\
\bottomrule
\end{tabular}
\end{table}
\subsection{Study Procedures}
\subsubsection{Phase 1: Deployment and Interviews}

Our study employed a multi-phase qualitative protocol that combined field deployment, patient interviews, ethnographic observation, and provider focus groups (see Fig. \ref{fig:studyphases}). Between July and August 2025, we introduced women 48 women to the chatbot during antenatal appointments. By September, only nine had engaged with it, and just five used it more than once. The overwhelming majority therefore did not use the system, foregrounding non-use which became central to our analysis. 

Initial interviews in July, conducted in crowded waiting rooms, were brief ($\approx$15 minutes) due to poor connectivity and constant interruptions. These covered demographics, technology access, and information sources, followed by chatbot onboarding. In remote follow-ups, only two to three responded, allowing for short semi-structured phone interviews ($\approx$15 minutes). 

In August, we adapted by expanding recruitment zones, reducing interruptions, and lengthening interviews ($\approx$45 minutes). Women briefly used the chatbot, then reflected on first impressions, trust, authority, and how digital use and advice might fit within their maternal health journey. Five additional in-depth interviews were conducted at a private clinic. Participants received a modest compensation of 1,000~PKR (\(\approx\)~US\$3.50). For context, Pakistan's monthly minimum wage of 32,000~PKR (\(\approx\)~US\$115).

Alongside interviews, we conducted ethnographic observation throughout the deployments, keeping detailed field notes on clinic environments, waiting room dynamics, interruptions, and family involvement. These notes became a critical resource for interpreting patterns of non-use and contextualizing women’s narratives.

\subsubsection{Phase 2: Provider Focus Groups}
Two focus groups ($\approx$1 hour each) with six gynecologists further contextualized patient perspectives on non-use. Discussions explored communication practices, perceived gaps in knowledge, responses to chatbots, and broader cultural and institutional factors shaping maternal care, including family involvement, gender norms, and systemic pressures.

\subsection{Data Collection and Analysis}
Most sessions were conducted in Urdu, audio-recorded, and later transcribed by the research team. Two unrecorded sessions were documented through detailed field notes. English translations were collaboratively produced to preserve cultural meanings of key terms (e.g., \textit{bardasht}, or endurance).

We conducted an inductive thematic analysis~\cite{charmaz2001grounded}, to remain closely grounded in participants’ accounts. We then contextualized these emergent themes within existing scholarship on maternal health, gender, and HCI research on chatbots, situating our findings in broader debates. Codes included relational authority, idioms of care, technology imaginaries, and institutional dynamics. Two researchers, with backgrounds in HCI, public health, and anthropology coded the data collaboratively, cross-checked themes, and resolved disagreements through discussion. Triangulation across patient interviews, provider focus groups, and field notes situated the chatbot within the maternal healthcare ecologies.

\subsection{Ethics and Researcher Positionality}
This study received Institutional Review Board approval, and oral informed consent was obtained from all participants, with explanations adapted to different literacy levels. Focus groups and interviews were recorded with permission. 

The first two authors were the interviewers, women from Lahore who shared Urdu as a first language with participants, which facilitated rapport and interpretation of idioms. At the same time, our institutional affiliations—as researchers from a private university connected through clinicians in tertiary hospitals—meant we were often perceived as educated, wealthier, and aligned with medical authority. Many participants, themselves low-income women, met us in spaces where they typically interacted with doctors, which likely amplified hierarchical dynamics. We sought to mitigate this by adopting an informal conversational style and building rapport beyond the interview protocol.

During analysis, two team members from different disciplinary backgrounds coded collaboratively, reflecting on perspectives shaped interpretation and refining themes iteratively.

One coauthor, a senior gynecologist with over 25 years of experience and head of a teaching hospital department, provided clinical expertise. This interdisciplinary collaboration allowed us to situate the chatbot as a technical system, as well as a social and clinical actor, grounding our analysis in the everyday realities of maternal healthcare in Lahore.

Our multi-phase qualitative design—combining field deployment, patient interviews, ethnographic observation, and provider focus groups—enabled us to study the chatbot not only as a digital health tool, but also as one situated within the institutional and relational landscapes of maternal healthcare in Lahore.

\subsection{Limitations}
This study does not claim representativeness across Pakistan. Participants were recruited through tertiary hospitals in Lahore, settings that already presuppose some degree of institutional access. Women who never reach such facilities, or who rely primarily on midwives and neighborhood clinics, are absent here. Even within hospitals, recruitment sometimes depended on doctor referrals, which may have filtered for women already more compliant or deferential. These dynamics inevitably shaped what participants shared and how they positioned us as university-affiliated researchers aligned with medical authority.

Nevertheless, the study offers an in-depth window into how relational authority, shared access, silence, and infrastructural fragility shape maternal health chatbot uptake in Lahore. The transferability of these insights lies less in statistical generalization than in their conceptual traction: these dynamics are likely to resonate across fragile health systems where pregnancy care is collective, phones are shared, and infrastructures remain uneven.

While many participants were low-income Punjabi women living in joint-family households and identifying as homemakers, there was meaningful variation in education, phone access, and information sources—factors that shaped engagement with the maternal health chatbot. Most participants resided in peri-urban areas of Lahore. 

\section{Findings}

Women’s engagement with the maternal health chatbot was never an individual act. It unfolded within family negotiations, clinical hierarchies, and fragile health systems. Across our fieldwork, five patterns stood out. Participant quotes are labeled P\#, clinician quotes D\#.

First, decision-making was mediated: consent and use were rarely individual but negotiated with husbands, mothers-in-law, and clinicians. Second, participation often took the form of endurance and silence, as women narrated discomfort or deferred questions rather than asking directly. Third, access was episodic, shaped by shared phones and intermittent availability, positioning the chatbot as a resource to consult-when-needed rather than a continuous companion. Fourth, women made sense of the chatbot through familiar social frames, likening it to a doctor or elder, with credibility tied as much to tone and voice as to informational accuracy. Finally, all of these practices unfolded within fragile conditions of maternal healthcare, where scarcity, bureaucratic overload, infrastructural breakdown, and climate disruptions made every encounter provisional. 

Together these patterns explain why use and non-use reflect relational authority and system fragility—setting up our RCDG in the discussion.

\begin{figure}
    \centering
    \includegraphics[width=1.0\linewidth]{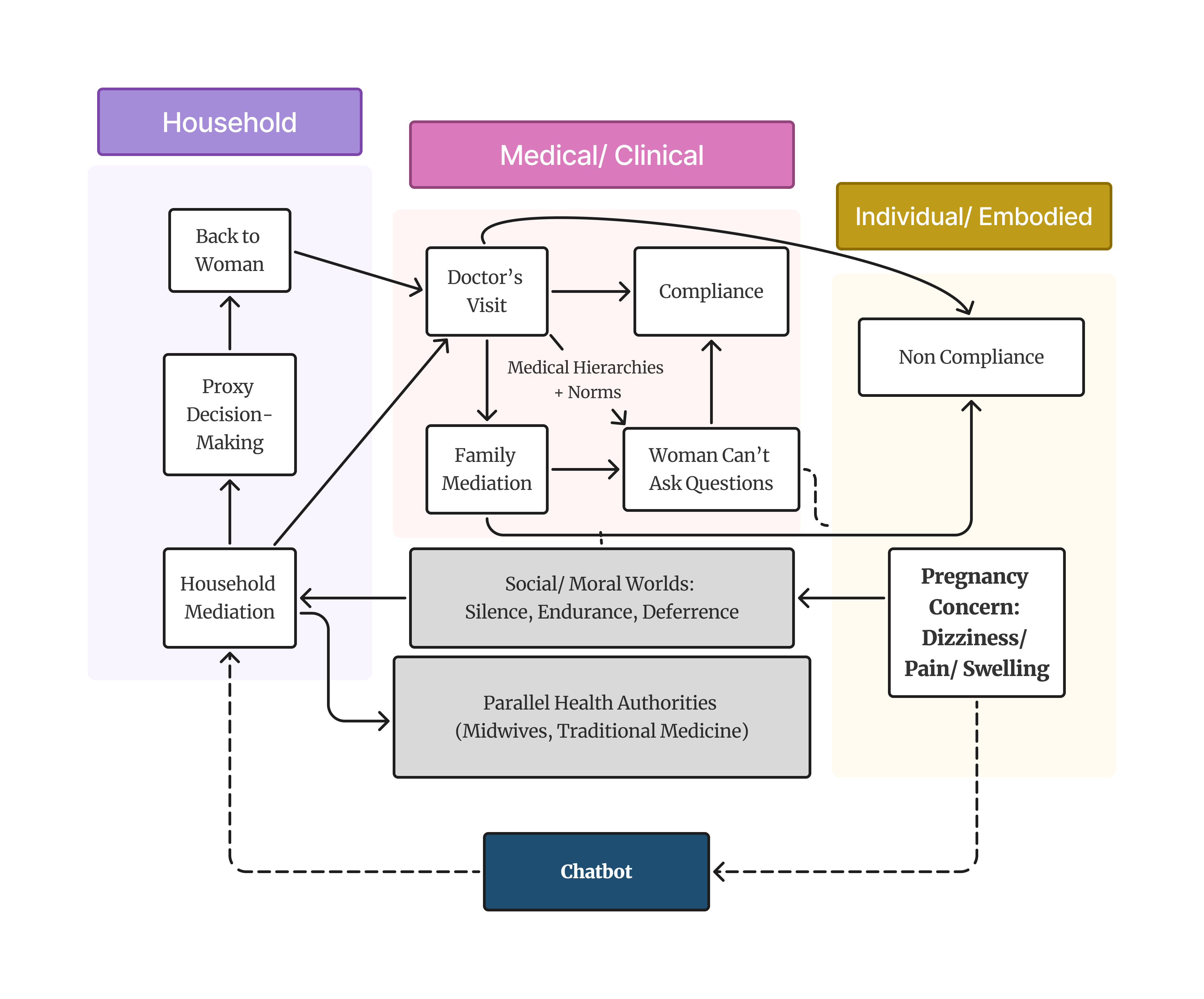}
    \caption{\textbf{ Pathways of pregnancy care negotiations.} Flow of symptoms, decision-making, and advice across individual, household, and clinical levels, mediated by family, social norms, and parallel health authorities. The chat bot enters this circulation; its guidance has to be legitimated or contested along with these existing negotiations.}
    \label{fig:pathways}
\end{figure}
\subsection{Mediated Decision-Making}

Consent to participate in the interview and to use the chatbot was rarely individual, but unfolded as a negotiation with family. A patient who initially agreed to an interview quickly added she must first consult her husband. Only after informing him did she return to say: \emph{``Yes, I can talk to you now.''} (P33, 25 years old). During onboarding, attendants occasionally intervened if they felt the field trial was taking too long. Many women requested that their mothers-in-law or sisters-in-law sit with them as they used the chatbot, often turning to them for affirmation, confirmation, or to recall details. In practice, the chatbot was rarely encountered alone: it was folded into collective presence and oversight, circulating across household, clinical, and individual domains (see Fig.~\ref{fig:pathways})

Clinicians framed this family presence as normal for maternal health care, but also as interference in both clinical and digital encounters. They described how attendants interrupted consultations, dismissing women’s complaints as exaggeration or \emph{''drama''} (D1). One doctor summarized: \emph{``Primigravidae have the least autonomy—often none''} (D4).
Among the primigravidae (women who are pregnant for the first time) we met (about 17 percent), none used the chatbot independently. One young participant (P34, 20 years old) who had studied up to grade 8, came from a low income background, and relied primarily on her mother-in-law for health advice—while occasionally turning to YouTube or Google—exemplified this tension: curious about the chatbot yet cautious about its consequences, she asked, \emph{“Sister, this won’t cause any problem in my household, right?”} When reassured that her chats would remain private and on her phone only, she replied: \emph{“No, I understand that, and I actually really liked it, but I was just asking whether whatever it says will have any effect on my household.”} Here, her hesitation illustrated how using the chatbot involved negotiating between personal curiosity and the authority of household hierarchies.

Another participant (P13, 20 years old), a pharmacist by profession, illustrated how women had to navigate family mediation against medical and digital authority. She recalled telling her mother and mother-in-law during her first pregnancy, at 19, that she was in pain, only to be told this was ordinary, or that she was making excuses to avoid work. \emph{“These days only doctors have the knowledge, no one else does…what the doctor knows, no one else does.”} After using the chatbot, she explained that she liked that the responses reflected this same clinical expertise, appreciating its trustworthy and personalized tone, \emph{“like a doctor.”} She indicated that she would like to continue using APPA in her second pregnancy, now, and added that it would be really useful to her if they bot would prescribe medication.

At the same time, many women described families as trusted conduits of care rather than obstacles. Some redirected authority deliberately, as recounted by doctors: \emph{“My mother-in-law or sister-in-law is here with me—you can explain it to them''} (D5). Others leaned on elder sisters and sisters-in-law for guidance: \emph{''She guides me the most, since she already has children. I also know a little from my first baby. Pregnancy is difficult, so you just have to be patient.''} (P9, 31 years old). For these women, the chatbot was not an isolated tool but something to be engaged with alongside kin who arranged transport, paid fees, accompanied them to clinics, and remembered instructions. Even when advice from their family was unwelcome—such as persistent encouragement to eat meat (P34, 20 years old)—women acknowledged its usefulness: they were, in their words, \emph{“taken care of.”}

In short, family mediation was both constraint and support. Prior work documents kin as gatekeepers and enablers in maternal health \cite{bagalkot2022embodied,vemireddy2021fragileinfra,batool2017maternal,sultana2018patriarchal,yadav2019feedpal,naveed2022privacyperceptions,mustafa2020patriarchy}. Our findings extend this to digital care: the chatbot was not an individual tool but part of household negotiations over knowledge, authority, and responsibility, reframing proxy decision-making as a constitutive condition of use rather than simply a barrier to uptake.

How can a maternal health chatbot work with family mediation as a source of support, while safeguarding women’s agency when it becomes a constraint? We return to these implications in the discussion section, situating them within broader debates on relational autonomy and non-use in HCI.

\subsection{Endurance and Silence as Participation}

Women’s engagement during live chatbot testing rarely took the form of direct questioning. Instead, they relied on practices such as withholding, deferring, narrating, and enduring to communicate their experiences of pregnancy to doctors, family members, the chatbot, and us—without presenting them as direct questions. These indirect modes of inquiry are rarely recognized in current chatbot design.

During onboarding, some women said they had \emph{``no questions for now,''} yet narrated dizziness, anemia, or persistent pain. For one interviewee, as we continued the conversation, she described that she simply had no questions during her pregnancy and did not turn to anyone for information or support, despite the fact that she had conceived for the first time after fourteen years of marriage, had described her pregnancy as precious, and was admitted to the hospital due to breathing issues (P04, 30 years old). Other interviewees posed their concerns not as interrogatives but as lived experiences—leaving facilitators and the chatbot itself to reframe narrations into actionable queries. Silence here was not absence of things to say in relation to their pregnancies and health, but participation through narration without conversion into explicit prompts.

For others, timing and context determined when a question "became askable": \emph{I just ask when I see the doctor''} (P30, 22 years old). Requests for rest, too, were withheld, framed as burdensome: \emph{``I am never a burden on anyone, nor have I bothered anyone by asking for things or saying `do this for me.' I just handle everything on my own.''} (P31, 37 years old). Here endurance itself became a communicative stance.

Endurance also structured how discomforts were normalized. On our last day at the hospital, we met P3, a 34-year-old teacher in her ninth month, admitted for gestational diabetes and struggling with breathlessness. One of the few who had even heard of ChatGPT, she explained she hadn’t sought help because she had believed that this was \emph{``just what a mother has to endure''} (P3, 34 years old).

While this is the most illustrative example of \textit{bardasht} or endurance, this motif came up across our engagement with women. Women frequently normalized pain, weakness, and emotional distress, thinking of these not as symptoms to be interrogated medically, but as conditions to be borne silently. Information, when encountered, often reinforced this stance. One participant described health videos from doctors appearing on her TikTok feed: \emph{"Their videos just come to me, but I’ve never followed anyone."} (P5, 28 years old). Here knowledge was absorbed passively, while others dismissed peers as overly sensitive:~\emph{"Some people get very sensitive. They say, I have this pain, I have that pain,' even though in this condition you will keep having pains here and there." }(P32, 30 years old).
Even psychological changes were narrated as fleeting states, not as problems requiring elaboration: \emph{``I don’t really remember in detail---just what I went through. Before, I didn’t feel much. Now it’s just the pain and things like that.''} (P38, 27 years old). And when complications such as miscarriage or neonatal death were mentioned, they were often framed as \textit{`Allah ki marzi'} (God’s will)—a religious stance that resisted framing such events as things to be considered medically at all.

Taken together, discomforts and uncertainties were narrated or deferred, but rarely posed as direct questions. This pattern has important implications for how engagement with chatbots should be understood.


\subsection{Episodic Use}
Phone access was rarely individual: even when women owned smartphones, these circulated within households, shaped by gendered routines in which husbands carried them outside, in-laws monitored use, and children often intruded on chats. Women sometimes requested mothers- or sisters-in-law to sit with them during chatbot sessions, underscoring that use was socially mediated. These dynamics transform the chatbot from an individual, personal health assistant into a household technology. Rather than serving as a private tool for health queries, the chatbot became a site of household medical discourse, with shared access shaping not only whether women could use it, but also how they imagined its authority and meaning.

Within these arrangements, women positioned the chatbot as valuable yet bounded. Some framed it as: \emph{``If I ever need any information, I can consult the app''} (P2, 22 years old). Such accounts underscored its role as a consult-when-needed resource, valued for reassurance but never a continuous companion.

Trust was conditional. One participant explained: \emph{``What the doctor says after seeing you is always the best. But if you want a second option\ldots then of course, you can just ask on a call or get some extra information.''} For others, even secondary reassurance was too risky: \emph{``What if the chatbot prescribed me a medicine that did not suit me? My doctor knows my full medical history, which is why I continued to follow only them.''} (P45, 20 years old).

Clinicians echoed this framing. They emphasized that patients often sought reassurance more than prescriptions: \emph{``What patients mostly need is reassurance---often, giving them peace of mind is more important than the medicine itself.''} (D5). Yet their own authority was also contested. Suspicion toward private hospitals complicated the hierarchy: \emph{``Private hospital doctors tend to go straight for a caesarean, even though they’re told they won’t be getting any extra payment for it.''} (D6).

Authority, then, was not binary---doctor vs. chatbot---but situated and partial, with both reassurance and suspicion shaping credibility across contexts. These accounts reveal how the chatbot became part of what we might call \emph{`household medical infrastructure'}---resources cataloged and remembered collectively for future need, like keeping a doctor's phone number or knowing which pharmacy stays open late.

These patterns of conditional and situated trust set the stage for how women made sense of the chatbot itself—not as an abstract AI system, but as something folded into familiar social worlds.

\subsection{Social Frames of Engagement: Making AI Legible}

Women's first reactions to the chatbot were often of curiosity and enthusiasm, imagining it as a resource that could extend beyond themselves. One participant reflected, \emph{“Had I known earlier, I would never have left it”} (P5, 28 years old). Others emphasized its usefulness for friends and family: \emph{“I will tell all my friends, so many are pregnant right now”} (P40, 25 years old), and ~\emph{“I already delivered, but I want to share it with my sisters-in-law who are pregnant; this is very useful.”} Another participant asked if it was also able to provide postpartum advice (P2, 22 years old). Such reactions underscored that women did not position the chatbot as an individual utility but as a collective resource that could circulate within families and social networks.

Participants rarely encountered AI as a technical abstraction. Instead, they domesticated its novelty through familiar social frames. One participant reflected: \emph{``It feels like someone is sitting right here beside me\ldots Yes, like a doctor.''} Another compared it to having an elder nearby: \emph{``It felt like there was an elder present.''} These were not mere metaphors but active processes of sense-making that attributed social presence to algorithmic responses.

Despite the chatbot's instant responses, some women who never tried it worried about delays: \emph{``But I was worried the reply might be delayed.''} (P45, 20 years old) This imaginative gap, where unfamiliarity with AI created anticipated barriers that did not necessarily exist, reveals how preconceptions about digital systems shaped engagement before any actual use. Women mapped their experiences with human medical providers (who might be busy or unavailable) onto the chatbot, not recognizing its fundamentally different temporal logic.

There was one case where a user was already familiar with, and used GPT for medical concerns: \emph{``I came here through the GPT app\ldots my blood pressure was very high, so it was an emergency\ldots they quickly performed a C-section.''} (P2, 22 years old). Here GPT functioned as a prompt for timely action, channeling women into clinical spaces.

This domestication, in turn, highlights how credibility was not only a matter of content or familiarity, but also of performance—particularly how authority was carried through voice.

Credibility in chatbot encounters also hinged not only on informational accuracy but also on sonic performance. One participant noted mispronunciations: \emph{``It’s just a small language issue\ldots like how it was saying `dooran.' I understood it, but some people might not''} (P10, 26 years old). The same woman described the robotic tone as both authoritative and distant: \emph{``It comes across as overly informative, as if no one is actually saying it themselves.''}

Clinicians similarly emphasized voice and tone as central to their own practice. In outpatient visits, they softened delivery to reassure: \emph{``In OPD cases, you can kindly reassure the patient that there’s nothing serious.''} (D1). In emergencies, they shifted to blunt authority: \emph{``But in a labor room emergency, the tone naturally becomes blunt---it can’t stay as soft.''} (D1).

Overall, women engaged with the chatbot through social imagination, often as a resource to share, a presence they could trust, or a voice they could trust or not trust. Credibility was shaped not just by technical accuracy but by how the system fit into existing social relations and performed authority in ways that made sense to them culturally.

\subsection{Clinicians’ Accounts of Compliance and Cultural Authority}

Although patients rarely mentioned~\textit\textit{dais}[unskilled midwives] or traditional remedies directly, doctors repeatedly emphasized how these forces shaped maternal care. These accounts reflected clinicians’ interpretations of the structural constraints that shaped compliance, as well as the competing cultural authorities they encountered in practice.  

\subsubsection{Compliance}
Clinicians saw compliance as being closely tied to both education and household resources. They noted that women with limited education tended to seek care only once complications had become urgent: \emph{“The villages around [the hospital in question] are linked to it, and people without education often arrive straight at the labor room emergency.”} (D2). By contrast, educated women were seen as maintaining long-term follow-up from the earliest stages of pregnancy: \emph{“They’re kept on follow-up from the point when the pregnancy test comes out positive but isn’t yet confirmed on the scan.”} (D5). Economic constraints were also part of these accounts. Families with several children, doctors explained, frequently prioritized school fees over medical visits: \emph{“Women often say, ‘We used the money for the children’s school fees instead of the hospital.’”} (D4).  

In these accounts, compliance was not portrayed as an individual choice but as structured by literacy, economic conditions, and competing demands on household resources.  

Beyond material factors, clinicians also emphasized gendered norms that dictated compliance. A senior consultant, and co-author on this paper, explained further: \textit{“No matter how much we talk about women’s liberation or awareness of rights, an ordinary homemaker here still feels her mind is suppressed—she is not tuned to ask questions for herself. Even educated women, graduates or postgraduates, have little autonomy in pregnancy. Their decisions are overwhelmingly influenced by family members—immediate or extended—who impose their opinions at every step. That is why even when a woman begins to ask a question, she often withdraws it herself.”} 

\subsubsection{Competing Systems of Care}
Doctors also positioned themselves against enduring parallel systems of care. They described how dais and local healers advised alternative remedies: \emph{“When we explain proper cord care to patients, they often respond, ‘No, our dai told us we should apply desi ghee mixed with stove soot.’”} (D3). Similarly, patients were said to seek reassurance from local clinics: \emph{“People without education often go to local clinics and feel satisfied with that.”} (D3).  

For clinicians, such practices represented both delays in medical intervention and rival systems of authority in maternal care. Cultural remedies and local healers were not dismissed as minor alternatives but cast as competing logics against which doctors had to assert medical expertise.  

\subsection{Fragile Conditions of Maternal Healthcare}

Before moving to the discussion, it is important to situate our findings within the systemic disruptions documented in our ethnographic notes and observations. For the purposes of our work, we define fragility as a systemic characteristic of maternal healthcare in Pakistan, where scarcity, institutional overload, infrastructural breakdown, and climate vulnerability combine to make care provisional and intermittent. Understanding chatbot use or non-use requires seeing it as a response within these structures rather than as a failure of adoption.

Even though the non-profit hospital we conducted the majority of our fieldwork at had clearer workflows and less visible chaos than a public hospital, women still felt that access to doctors was precarious.  

In such a system, even small lapses of attention could mean losing access to care that was already hard to secure. At our field site, women sat through long hours focused only on not missing their turn, since a single consultation often took an entire day at the hospital~\cite{bagalkot2022embodied}. For patients like P30, who saved up all her questions for that brief encounter, missing the appointment meant losing the doctor’s check-up and advice entirely. Women’s anxiety reflected not just one hospital’s workflow but the broader scarcity of care. In this context, Patient 31’s insistence that she would \textit{`never be a burden to anyone'}, shared earlier, can be read not only as cultural deference but also as a way of coping with an environment where care was intermittent, difficult to access, and quickly exhausted.

Clinicians, too, were pressed for time—a pattern not unique to this hospital but part of wider health system shortages. Each consultation became a compressed, high-stakes encounter where women tried to prepare and absorb everything at once. Our earlier findings, where women preferred to have their families or attendants with them (P9 and P34, for example), also show how they managed this pressure. Family members remembered advice, asked questions that women missed, and clarified when needed. The burden of scarcity was shared: doctors worked under constant overload, while women relied on family networks to distribute the cognitive and emotional load of navigating fragile healthcare.

Fragility revealed itself not only in resource shortages, but in the system-level frictions and institutional overload that forced care to be continually patched, improvised, and renegotiated. One day, a senior doctor we knew rushed past with stacks of papers—not patient charts, but applications for a delayed monthly allowance. The stipend required fresh paperwork each month, with no system in place. These bureaucratic struggles layered onto clinical overload, illustrating how fragility drains time and attention away from patient care.

Technical infrastructures compounded this precarity. Within the hospital, we found that signals were blocked in the outpatient department where we were working and needed stable WiFi for the chatbot to be live tested by interviewees. When we asked why, an administrator responded with a shrug: \textit{“This is Pakistan—our staff get distracted by social media.”} In practice, what mattered for us was that connectivity could not be taken for granted.

During our weeks of fieldwork, Punjab was struck by some of the heaviest monsoon rains and worst urban flooding in years. Attendance at the hospital fluctuated with the heavy monsoon rains, underscoring how fragile infrastructure directly constrained access to care. Vulnerabilities in transport and drainage extended well beyond the hospital walls, turning seasonal weather into a structural barrier for both patients and providers.

These disruptions—climate crisis, policy neglect, and chronic under-resourcing of reproductive health—were not incidental, but the broader structures within which women encountered the chatbot. In the discussion, we frame this as a call to design with fragility as the baseline, rather than assuming stability. They are constitutive elements of the sociotechnical systems within which any HCI intervention must function.

\begin{figure}
    \centering
\includegraphics[width=0.6\linewidth]{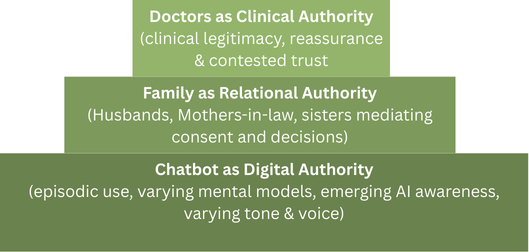}
    \caption{\textbf{Pyramid of relational authority in maternal health chatbot use.}
Doctors anchored authority, kin mediated consent, and chatbots were situational and relationally framed, showing adoption as collective rather than individual.}
    \label{fig:pyramid}
\end{figure}
Taken together, these findings reveal that  maternal health chatbot engagement is organized through a pyramid of hierarchies (Figure \ref{fig:pyramid}): doctors anchored legitimacy, family mediated consent and access, and chatbots were positioned as supplemental and situational. Silence and deference structured when and how questions were asked; shared devices and family presence shaped availability; and credibility hinged on tone as much as on content. Clinicians’ accounts located compliance in education and household resources and cast cultural remedies as competing systems of care. The discussion section builds on these observations.

\section{Discussion}

\begin{figure}
    \centering
    \includegraphics[width=0.9\linewidth]{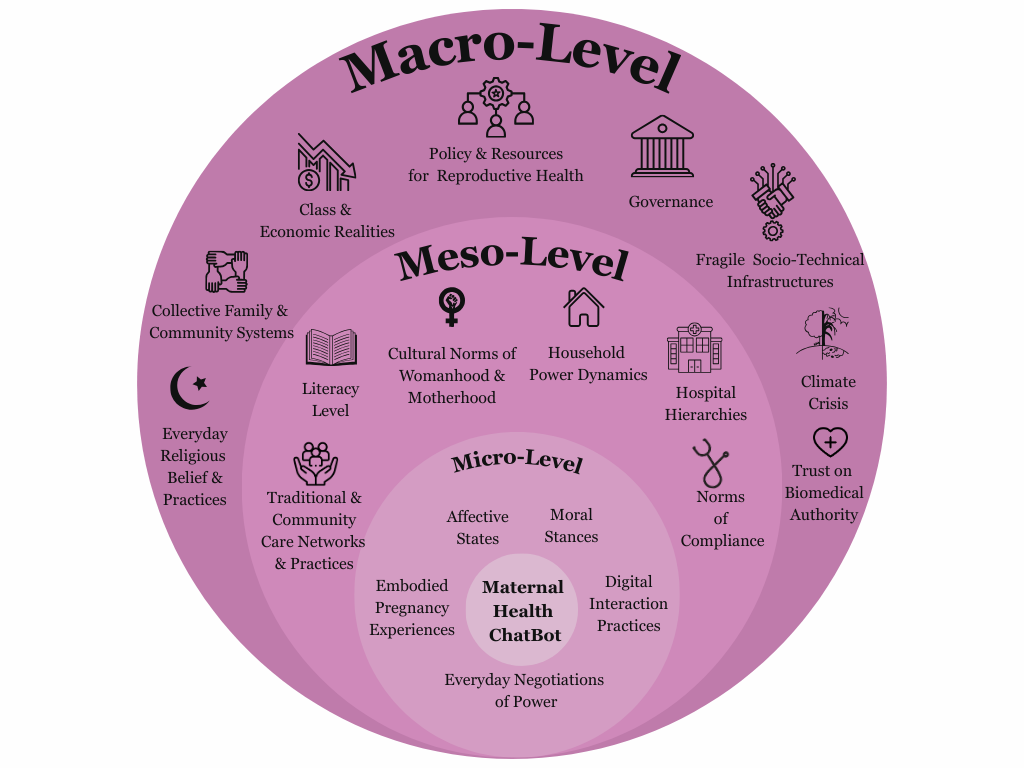}
    \caption{\textbf{Nested dynamics of maternal health chatbot use.} The figure illustrates \textit{micro-level} embodied experiences, \textit{meso-level} household dynamics, and \textit{macro-level} structural conditions intersect to shape how women engage with chatbots. This ecological view grounds our Relational Chatbot Design Grammar, framing silence, family mediation, episodic use, and fragility as key design commitments.}
    \label{fig:nested dynamics}
\end{figure}

Our findings reveal that maternal-health chatbot use is shaped by family mediation, clinical authority, episodic access, and routine breakdowns. Building on these observations, we contribute a \textit{Relational Chatbot Design Grammar (RCDG)}---a set of four commitments that reframe maternal health chatbot design in fragile, collective care contexts (see Fig.~\ref{fig:nested dynamics}).

\begin{itemize}
    \item \textbf{Mediated Decision-Making:} Designing for maternal health means treating authority as relational—mediated through kin and clinicians—so chatbots work with, rather than against, these negotiations.
    
    \item \textbf{Silence and Endurance:} Chatbot design needs to recognize silence, withholding, and narration as active ways women engage with care, beyond explicit queries.
    
    \item \textbf{Episodic Use:} Chatbot design should account for episodic engagement, valued for reassurance and consultation in moments of need rather than as continuous companionship.
    
    \item \textbf{Fragile Contexts:} Breakdowns in networks, institutions, and households are everyday conditions; design must assume disruption and plan for survivability.
\end{itemize}

In what follows, we unpack each commitment, situating it in relation to prior CHI work and showing how it sets up the design recommendations that follow.
\subsection{How Do Chatbots Navigate Mediated Decision-Making?}
This illustrates our first RCDG commitment, design for mediated decision-making.

Even context-attentive work—such as ASHABot in India~\cite{khullar2025ashabot} or culturally sensitive family-planning chatbots~\cite{deva2025culturalsensitivity}—still tends to foreground translation, literacy, or question framing as the primary adaptations to localize and contextualize chatbots. By contrast, our findings point to something deeper: in Pakistan, reproductive life is collective, with consent and legitimacy negotiated within household and clinical hierarchies. Our interviews with clinicians confirmed this, recalling how even highly educated women were heavily influenced by the opinions of their family members.

A social justice–oriented maternal health chatbot must therefore move beyond Western notions of individual consent, focusing instead on sustaining women’s voices within family and clinical negotiations. If women are to adopt chatbots as companions in their pregnancy journeys, as many interviewees expressed a desire to do, design must help carry their narrations, hesitations, and needs into spaces where they risk being silenced. We expand on this with examples in our design commitments.

To center feminist scholarship from the Global South, empowerment and agency are relational, recognized and sustained through collective worlds~\cite{Kabeer1999,rowlands1997questioningempowerment,abulughod1990veiled}. Deference or proxy decision-making may be strategies for navigating power by gaining recognition and support within families and communities, not just failures of autonomy and independence. These findings push us to resist designing chatbots that only encourage more queries from women. Instead, they call for attending more closely to how women already navigate power and for designing technologies that work with these relational practices rather than against them.

The design question, then, is when to align with existing trust relations and when to push against them to protect the safety and dignity of the individual woman. How do we build maternal health chatbots that amplify women’s capacity to negotiate, be recognized, and act collectively within these layered structures of care?

\subsection{ How Can Silence and Endurance Be Recognized as Engagement?}

If shared authority highlights who speaks, silence reveals how participation is enacted. This subsection reflects our second RCDG commitment, silence and endurance as engagement.

Women withheld, deferred, narrated, and endured in ways shaped by family, religion, and clinic. These practices, though rarely direct queries, structured how pregnancy was communicated. Instead, they press us to reconsider what counts as participation in maternal health technologies, and to ask how systems might accommodate communicative forms that emerge through narration, endurance, and deference rather than explicit questioning. A co-author, drawing on decades of clinical practice, recalled that many women describe discomforts but rarely frame them as inquiries; what they seek is reassurance and continuity, not answers to questions they have never been taught to pose.

Doctors noted that women were not attuned to ask questions for themselves, while hospital hierarchies reinforced this dynamic.

To read silence as disengagement is to miss its meaning. Feminist scholars such as Mahmood~\cite{mahmood2005politicsofpiety} and White~\cite{white2010fundamentalcauses} have shown us how silence and deference can be active moral stances, means of sustaining dignity and belonging. Endurance and perseverance, then, is not the failure to participate but a particular way of communicating care.

Current technology paradigms, especially in health, overvalue speech—countable queries and explicit prompts—and overlook communicative forms grounded in endurance~\cite{xu2017nonuse,progga2025perinataltech,silva2024chatdesign,liu2018chatbots}. When engagement is reduced to questioning, systems privilege those who ask and erase endurance-based practices.
The design challenge is not to extract more queries but to recognize diverse communicative forms. Systems should scaffold narration, recognize endurance as communicative, and allow discomforts to be translated into actionable prompts without requiring women to ask in unfamiliar ways. This calls for interaction patterns (narration-to-question translation, episodic check-ins, recognition prompts) and evaluation metrics that credit narrated distress as participation.

While decision-making authority reveals \textit{who} is permitted to speak and act, our second finding—silence—shows that even when women are allowed to participate, \textit{how} they communicate care is structured by endurance and deference. Together, these commitments push us to see both authority and communicative form as relational, rather than reducing engagement to an individual woman’s willingness to ask questions.

\subsection{What Does Episodic Use Reveal About Trust in Maternal Health Chatbots?}

This subsection brings together two RCDG commitments: Mediated Decision-making and Episodic Use. 

Our findings show that trust is never located in the chatbot itself but was continually negotiated across doctors, families, and institutions. Women consistently deferred to doctors while suspicion toward private hospitals complicated biomedical legitimacy. This aligns with Ismail and Kumar’s critique that AI-for-health systems often ignore governance and authority~\cite{ismail2021frontlines}, and with Meyrowitsch et al., who caution that AI can amplify misinformation among vulnerable groups~\cite{meyrowitsch2023misinfo}. It also diverges from chatbot studies that equate trust with accuracy, transparency, or empathetic tone~\cite{silva2024chatdesign,liu2018chatbots,progga2025perinataltech}. In our context, trust was relational infrastructure—grounded in clinical authority, family endorsement, and institutional legitimacy—echoing but extending findings from LMIC deployments such as ASHABot, where bots were trusted when aligned with guidelines~\cite{deva2025culturalsensitivity}. 

Trust was also shaped by the episodic nature of engagement. While HCI4D research acknowledges infrastructural limits~\cite{verdezoto2021maintenance,coleman2023reconsidering,till2023digitalhealthafrica}, it often assumes individualized, continuous access once WhatsApp is available. Our data complicates this: phones were shared among husbands, in-laws, and children; even when women owned devices, they were often left at home when they came to the clinic. Engagement was therefore intermittent, with the chatbot positioned as a reference to consult-when-needed, not a companion carried through pregnancy. This reframes engagement from persistence to resonance, resonating with Wong-Villacres et al. on asset-based design~\cite{wongvillacres2020assetbased} and Bagalkot et al. on relational support~\cite{bagalkot2020beyondliteracy}. Yet chatbot research rarely evaluates episodic use, still privileging query counts as indicators of uptake. Designing for intermittent access means building redundancy (voice, SMS, WhatsApp) and recognizing that family, along with individuals, may carry the technology forward. 

A final dimension lies in tone as a performance of authority. Conversational design often assumes that naturalness, empathy, and human-like style increase trust~\cite{silva2024chatdesign,liu2018chatbots}. Yet in clinical practice, tone is not neutral—it is how authority and safety are enacted. Doctors described softening tone in outpatient consultations but shifting to blunt, directive urgency in emergencies, a tonal pivot patients trusted as a signal of seriousness. By contrast, the chatbot’s robotic mispronunciations undermined comprehension but paradoxically added authority, sounding “too informative,” as if beyond human partiality. 

This complicates assumptions that human-likeness uniformly increases trust. Instead, authority was enacted through knowing when and how to signal urgency and guide action in ways that were recognized as caring and legitimate. This insight extends Henry et al.’s feminist design justice approach~\cite{henry2024designjustice}: technologies must embed responsiveness to context. For health chatbots, this requires directive emergency modes—communicative pivots from reassurance to urgency, backed by human-reviewed escalation protocols. In fragile systems, safety depends not on empathy alone but on embedding authority that can survive shared phones, noisy clinics, and familial relay.

\subsection{How Can Fragility Become a Baseline for Design?}
This subsection reflects our fourth RCDG commitment, Fragile Contexts. 

In our field sites, connectivity faltered, bureaucratic suspicion derailed access, staff turnover disrupted continuity, and women frequently moved between households late in pregnancy. Systematic reviews have catalogued these as “barriers”~\cite{hussain2025revolutionize,fuentes2024emerginghai,till2023digitalhealthafrica}. Verdezoto et al. foreground the invisible maintenance labor sustaining community health systems~\cite{verdezoto2021maintenance}, and Coleman et al. show how South African women reframe digital health around continuity and trust~\cite{coleman2023reconsidering}. Yet chatbot design often assumes stability, focusing on usability while overlooking the instability that structures this use. 

Designing for fragility requires building redundancy (multi-channel delivery), distributing legitimacy across clinicians and family, and ensuring resilience to institutional flux. Does a red flag raised from conversation still reach the user when Wi-Fi is intermittent? Does the advice remain credible when hospital politics shift? This also shifts evaluation: the key question is not whether systems “scale,” but whether they survive breakdowns. 

This also means that evaluation metrics should value resilience, relational uptake, and survival across disruptions rather than  continuous, individualized engagement alone. Building on prior LMIC chatbot deployments~\cite{montenegro2022pregnancychatbot,deva2025culturalsensitivity,progga2025perinataltech}, feminist and HCI4D critiques~\cite{sultana2018patriarchal,wongvillacres2020assetbased,henry2024designjustice}, and systematic reviews~\cite{hussain2025revolutionize,fuentes2024emerginghai,till2023digitalhealthafrica}, but diverges by treating silence, family authority, and fragility as conditions of possibility rather than deficits.

Unless chatbots account for these structural realities, maternal-health systems will continue to see them as technical curiosities with high drop-off and low trust. Our contribution is to show empirically how fragility operates as a baseline in Lahore and to articulate a design grammar for relational, fragile contexts—a lens through which global HCI can rethink conversational agents for health, well beyond Pakistan.

\section{Design Recommendations}

We present the Relational Chatbot Design Grammar (RCDG), which reframes maternal health chatbot design around four commitments: Mediated Decision-Making, Silence and Endurance, Episodic Use, and Fragile Contexts. We also consider the transferability---the applicability of these commitments beyond maternal health, positioning them as design orientations that can inform technologies in other domains where care, authority, and fragility are central conditions of use.

Taken together, these raise critical questions about maternal chatbots in fragile, collective care contexts:
\begin{itemize}

    \item How can systems support protective silence and withdrawal instead of equating engagement with empowerment?
    \item How can we build digital health tools that tell the difference between strategic non-use and forced absence—and measure success by moments of need rather than constant use?
    \item How can we design technologies that help people survive in neglected systems, opening room for dignity instead of mere adaptation or compliance?
    \item When does collective care risk turning into monitoring, and how can technologies recognize this to protect users while still offering support and tactical ways to face this mode of control?
\end{itemize}
\subsection{Designing for Mediated Decision-Making}

Consent and chatbot use were shaped not by individuals but by family hierarchies and clinical authority, requiring systems that support collective decision-making without disrupting it. 

\begin{itemize}
    \item \textbf{Enable layered consent:} Systems should let women decide which actors to involve and when, sustaining agency through selective sharing.
    
    \begin{itemize}
        \item \textit{Local example:} A woman may want iron supplement reminders shared with her husband; the chatbot allows her to share the reminder with him as well.  
        \item \textit{Transferability:}  Layered consent helps chatbots fit into collective care contexts while preserving autonomy.
    \end{itemize}

    \item \textbf{Preserve women’s conversations:} With consent, the chatbot should pass on what women say in their own words to their clinicians, not rephrased.  

    \begin{itemize}
        \item \textit{Local example:} If a woman says, ``I feel dizzy,'' the chatbot escalates: ``The patient reports, `I feel dizzy since yesterday.' Please review.''
        \item \textit{Transferability:}  Symptom based escalation that preserves the woman's concern help women's concerns to be prioritized in collective care contexts while preserving autonomy.
    \end{itemize}
\end{itemize}

\subsection{Designing for Endurance and Silence}

Designing for silence as engagement means recognizing quietness and indirect narration as valid forms of engagement. 

\begin{itemize}
    \item \textbf{Treat indirect expressions as meaningful:} Instead of discarding vague or minimal responses, the chatbot should reframe them into supportive statements, guidance, or optional summaries.

        \begin{itemize}
            \item \textit{Local example 1:} If a woman types “sometimes tired,” rather than asking diagnostic questions, the system could reply: “Tiredness during pregnancy is common. Would you like gentle ways to manage energy? I can save this privately or create a family-friendly summary if that would help.”
            \item \textit{Local Example 2:} If a woman repeatedly gives short replies about mood (“okay,” “fine,” “same”), the system might say: “I notice you’re keeping things brief. That’s fine. Would you like quiet check-ins or resources you can access when ready? I won’t save these conversations unless you tell me to.”

            \item \textit{Transferability:}  In contexts where deferring or understating is common, such reframing ensures quiet expressions are not dismissed and remain meaningful contributions to care.
        \end{itemize}
          
    \item \textbf{Invite non-verbal engagement:}  While most chatbot frameworks are optimized for continuous text, they can be adapted to register minimal inputs (emoji reactions, single-character taps, timed non-responses) as legitimate engagement~\cite{silva2024chatdesign,jo2023publichealthllm}. 

        \begin{itemize}
            \item \textit{Local example:} After an appointment, the chatbot might ask, “Did you go for your check-up this week?” with one-tap responses such as “Yes,” “No,” or “Planning to”. 
            \item \textit{Transferability:} Designing for episodic and non-verbal engagement keeps technologies usable in irregular access contexts, where short acknowledgments are more realistic than continuous interaction.
        \end{itemize}
        
\end{itemize}

\subsection{Designing for Episodic Engagement}

Rather than treating episodic use as a user experience problem to solve, we must recognize it as both imposed constraint, sometimes, a tactical response to forms of digital surveillance and control. 

\begin{itemize}
    \item \textbf{Use simple, trustworthy language}. Systems can use clear, locally rooted language that families can relate to, offer “side-by-side advice” contrasting common remedies with evidence-based guidance, and state their limits openly so expectations stay realistic.

        \begin{itemize}
            \item \textit{Local example:} A woman returning after weeks receives a short message about trimester changes, phrased in familiar terms, and simple enough to repeat and share with her mother-in-law. This also includes a reminder about her next antenatal appointment would be in this week and it is important that she attends it.
            \item \textit{Transferability:} Communication that is culturally resonant, repeatable, and transparent enables digital systems in fragile contexts to be trusted and acted upon even when access is irregular.

        \end{itemize}  

    \item \textbf{Rethink evaluation beyond frequency.}  Many mHealth evaluations overwhelmingly rely on engagement success with sustained engagement—daily log-ins, time spent, or number of sessions~\cite{wang2022snehai,mcmahon2023nursenisa,yadav2019feedpal,rahman2021adolescentbot,montenegro2022pregnancychatbot}. A maternal health chatbot may be better judged by whether it provides clarity and reassurance at the moments that matter, even if those moments are rare.

        \begin{itemize}
            \item \textit{Local example:} A woman may consult the chatbot only at the start of each trimester to understand expected changes. Though she uses it just three times, each interaction provides clarity and reassurance at a critical moment.
            \item \textit{Transferability:} In episodic contexts, evaluation must move beyond traditional engagement metrics. Design context-specific metrics and ways to capture them post-intervention that acknowledge the realities of intermittent use.

        \end{itemize}

    \item \textbf{Contextual adaptation.} Systems can be designed to notice when a woman’s conversational style changes suddenly and adjust settings—like privacy—without requiring her to ask. Using natural language processing, the chatbot can detect these shifts and adapt how it responds.

        \begin{itemize}
            \item \textit{Local Example:} A user who previously asked straightforward questions like "What foods are safe during pregnancy?" returns with queries such as: "What if someone can't go to appointments?" or "How do you know if stress is too much?" The system detects this semantic shift and responds: "I notice you're exploring different topics now. I can adjust how I work—would you prefer if our conversations were automatically deleted after each session, or would you like other privacy options?"
            \item \textit{Transferability:} Across fragile or collective care contexts, adaptive systems that adjust privacy and response styles can ease the demand on users to signal every change.
        \end{itemize}

\end{itemize}

\subsection{Designing for Fragile and Disruptive Contexts}

Designing for fragility as normal means preparing for disruption, survival, and re-entry rather than assuming stability.

\begin{itemize}
    \item \textbf{Build redundancy across channels.}  Systems should ensure that essential information can travel through multiple modalities so that breakdowns in one do not cut off access entirely.

        \begin{itemize}
            \item \textit{Local example:} When WhatsApp messages fail during connectivity breakdowns, SMS or voice calls can still ensure urgent updates are delivered.
            \item \textit{Transferability:} Redundancy is critical in fragile infrastructures ensuring information persists despite unreliable connectivity or platforms.

        \end{itemize}

    \item \textbf{Support seamless re-entry.} Systems should allow users to pause and return seamlessly, providing continuity even after long interruptions caused by structural factors. 

        \begin{itemize}
            \item \textit{Local example:} A woman who stopped using the chatbot during her husband's unemployment, when phone credit became expensive and device sharing increased, was able to return a month or two later. The system recognized her previous concerns about low iron levels and offered: "Welcome back. I remember you were managing iron deficiency. Would you like a quick update on what to expect at 28 weeks, or do you have new concerns to discuss first?" rather than starting over or asking questions about her absence.
            \item \textit{Transferability:} In contexts disrupted by migration, conflict, or instability, re-entry flows prevent exclusion and help users rebuild continuity after gaps.

        \end{itemize}

    \item \textbf{Integrate disruption-aware check-ins.} Systems can acknowledge ongoing crises and adapt advice accordingly, reframing disruption as part of the user context rather than an aberration.

    \begin{itemize}
        \item \textit{Local example:} When severe flooding or political protests disrupted access to clinics, the chatbot can prompt women to mark themselves safe and suggested alternate routes, interim practices for care, and alternate resources for urgent needs.
        \item \textit{Transferability:} Disruption-aware check-ins extend to fragile contexts shaped by disasters, protests, or institutional churn, allowing digital systems to remain relevant when routine infrastructures falter.
  \end{itemize}
\end{itemize}

This design grammar offers a blueprint for technologies that take fragility, silence, episodic use, and proxy decision-making seriously—asking HCI not how to overcome them, but how to build systems that can work within them.

\section{Conclusion}

This study shows how women and families in Lahore engaged with a maternal health chatbot within the routines of pregnancy care. We found that engagement was negotiated through family and clinicians, enacted through silence and endurance, and constrained by episodic access on shared phones, all of which unfolded under fragile structures of healthcare. 

Rather than treating these dynamics as barriers, we frame them as baseline conditions of design. Family mediation, silence, episodic use, and fragility are not exceptions to be overcome but the everyday ground realities on which maternal health technologies operate. To capture this, we articulated a Relational Chatbot Design Grammar (RCDG) with four commitments: designing for mediated decision-making, recognizing silence and endurance as engagement, supporting episodic use, and assuming fragility as normal.

We center non-use as a patterned, relational practice rather than mere attrition.  expected modes of engagement and designing for them through these commitments. By embedding design within relational authority and fragile infrastructures, we show how chatbots should be reimagined as technologies that live within collective and precarious systems of care.

\bibliographystyle{ACM-Reference-Format}
\bibliography{main}
\end{document}